\documentclass[
  ,draft            % use draft while you are working on the paper
  ]
  {aipproc}

\layoutstyle{8x11double}
\usepackage{journals}

\begin{document}

\title{Do the low PN velocity dispersions around elliptical galaxies imply
  that these lack dark matter?}


\author{Gary A. Mamon}{
  address={Institut d'Astrophysique de Paris (UMR 8111: CNRS \& Univ. P. \&
M. Curie), Paris, France}
}

\author{Avishai Dekel}{
  address={Racah Institute of Physics, Hebrew Unversity, Jerusalem, Israel}
}

\author{Felix Stoehr}{
  address={Institut d'Astrophysique de Paris (UMR 8111: CNRS \& Univ. P. \&
M. Curie), Paris, France},
% \altaddress={Laboratoire d'Oc\'eanographie Dynamique et de Climatologie (UMR
%   7617: CNRS, IRD \& Univ. P. \& M. Curie, Paris, France}
}

% \author{Ewa {\L}okas}{
%   address={Copernicus Astronomical Institute, Warsaw, Poland}
% }

\begin{abstract}
While kinematical modelling of the low PN velocity dispersions observed
in the outer regions of elliptical galaxies
suggest a lack of dark matter around these galaxies, we report on an analysis
of a suite of $N$-body simulations (with gas) 
of major mergers of spiral galaxies
embedded in dark matter halos,  and find that the outer velocity dispersions
are as low as observed for the PNe. The inconsistency between our dynamical
modelling and previous kinematical modelling is caused by very radial stellar
orbits and projection effects when viewing face-on oblate ellipticals.
Our simulations (weakly) suggest the youth of PNe around
ellipticals, and we propose that the universality of the PN luminosity
function may be explained if the bright PNe in ellipticals are formed after
the regular accretion of very low mass gas-rich galaxies.

\end{abstract}

\maketitle

%%%%%%%%%%%%%%%%%%%%%%%%%%%%%%%%%%%%%%%%%%%%
%% MAINMATTER
%%%%%%%%%%%%%%%%%%%%%%%%%%%%%%%%%%%%%%%%%%%%

\section{Introduction}

Although it is generally accepted that spiral galaxies are embedded in dark
matter 
%(hereafter DM)
 halos, the presence and amount of dark matter in elliptical
galaxies is still uncertain.

The classical way to infer the mass distribution of a gravitational system is
by analyzing its internal kinematics. For an unordered system such as an
elliptical galaxy, the more mass there is, the higher the velocity
dispersions.
Mathematically, this is expressed through the Jeans equation 
$\nabla P = -\nu \,\nabla \Phi$, where $P$, $\Phi$ and $\nu$ are the
dynamical pressure, gravitational potential and (number or luminosity)
density of the tracer, which in spherical symmetry becomes:
\begin{equation}
{d\left (\nu \sigma_r^2 \right) \over dr} +
2\,\beta(r)\,{\nu(r)\,\sigma_r^2(r)\over r} 
= -\nu\,{G\,M(r)\over r^2} \ ,
\label{jeanseq}
\end{equation}
where $M$ is the mass, $\sigma_r$ the radial velocity
dispersion, $\beta = 1- \sigma_t^2/(2\,\sigma_r^2)$ the velocity anisotropy
(zero for isotropic systems) and $\sigma_t$ the tangential velocity dispersion. 
The modelling is complicated by projection effects and by the intrinsic
mass/anisotropy degeneracy in the Jeans equation.

Alas, stellar velocity dispersions in elliptical galaxies are difficult to
measure (through absorption-line spectroscopy) beyond 2 half-light
(\emph{effective}) radii ($2\,R_e$).
Planetary nebulae (hereafter, PNe) happen to provide a discrete indicator of
the gravitational potential that extends up to $5\,R_e$, and therefore helps
constrain the mass profiles of ellipticals.
Measured velocity dispersions of PNe around ellipticals turn out to be low
and decreasing with radius
(\citealp{CJD93,Mendez+01,Romanowsky+03}, but see Teodorescu, in these
proceedings), and refined kinematical
analyses lead to 
a lack of dark matter around elliptical galaxies (\citeauthor{Romanowsky+03}).
This conclusion is at odds with the predictions of the standard hierarchical
scenario in which elliptical galaxies are the products of mergers of spiral
galaxies of comparable mass 
%\citep{Mamon92,SWTK01}.
\citep{SWTK01}.
Do the low PN velocity dispersions force us to revise our general
understanding of the formation of elliptical galaxies?
In this contribution, we gain considerable insight through an analysis of 
dynamical $N$-body
simulations of galaxy mergers.

\section{The line-of-sight velocity dispersion dispersion profiles of merger
remnants}

We use a series of 10 simulations \citep{CJPS05} of equal mass mergers of
disk+bulge+gas+dark 
matter halo spiral galaxies, with different orbital parameters, bulge/disk
ratio, 
%initial 
gas fraction and feedback efficiency.
\begin{figure}
  \includegraphics[width=\hsize]{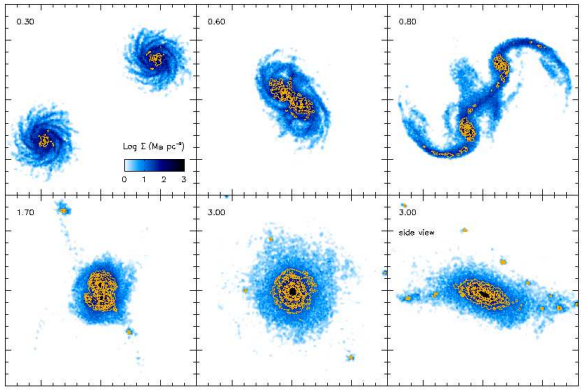}
  \caption{Snapshots of the merger of equal mass spiral galaxies (see
  \citealp{CJPS05}).  The lighter
zones in the central regions of the galaxies represents the young stars
  formed during the simulation, while the remaining smooth greyscale are the old
  stars.\label{snapshots}}
\end{figure}

Figure~\ref{snapshots} shows snapshots of one of the mergers, 
illustrating
the tidal tails and bridge formed during the closest approach, after which
the galaxies separate and merge at second passage, while the
tidal tails expand outwards. The final remnant, viewed here edge-on is oblate.

\begin{figure}
    \includegraphics[width=\hsize]{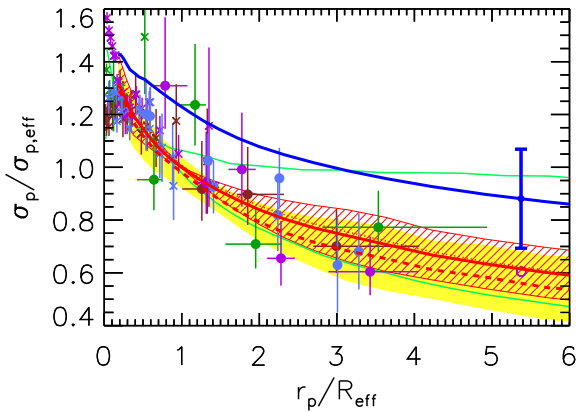}
 \caption{Line-of-sight
   velocity 
   dispersion vs. projected radius, normalized to the effective radius 
(from \citealp{Dekel+05}).
The merger remnants are viewed
from three orthogonal directions at two late times after the merger.
\emph{Lower solid}, \emph{dashed}, and \emph{upper solid curves} show the
 dispersions from all stars, young ($<3$ Gyr) stars, and dark matter, respectively.
The $1\sigma$ scatter 
is marked by \emph{shaded} or \emph{hashed areas} or a \emph{thick bar}.
For comparison, 
are shown 
%as \emph{green}, \emph{violet}, \emph{brown}, and \emph{blue}, 
the stellar (\emph{crosses}, \citealp{JS89,BDI90,BSG94,SS99}) 
and PN (\emph{circles}, \citeauthor{Mendez+01,Romanowsky+03}) 
velocity dispersions of 
galaxies NGC~821, 3379, 4494 and 4697.
The errors are $1\,\sigma$.
\emph{Thin curves} refer to 
the earlier models (\citeauthor{Romanowsky+03}) with (\emph{upper}) and without
(\emph{lower}) dark matter.
\label{siglossims}}
\end{figure}

Figure~\ref{siglossims} 
shows the line-of-sight velocity dispersion of the 10
merger remnants, together with the stellar and PNe dispersions of
\citeauthor{Romanowsky+03} and \citeauthor{Mendez+01}.
Interestingly, the stellar line-of-sight velocity dispersions of the dark
matter-embedded merger
remnants are as low as observed, and surprisingly, consistent with the
predictions of no dark matter of \citeauthor{Romanowsky+03}.
Also, the surface density profiles of the simulated merger remnants 
match almost perfectly the observed
surface brightness profiles \citep{Dekel+05}.
In other words, the simulated merger remnants match the observations, while
they are formed through the mergers of spiral galaxies embedded in dark
matter halos, which is in contradiction with the results 
from kinematical modelling (\citeauthor{Romanowsky+03}) that point to a lack
of dark 
matter in elliptical galaxies.

\section{Why did kinematical and dynamical  modelling produce inconsistent
results?}

One can criticize the relevance of these simulations to understanding the
kinematics of PNe around elliptical galaxies as these are not thought to be
subject to
recent (few Gyr) mergers of equal mass spiral galaxies.
However, the low velocity dispersions are also found in mergers of spirals
with mass ratio 3 (which are much more frequent) and these low velocity
dispersions are long lived if one lets
the remnant evolve for many Gyr more.

If the merger simulations analyzed by \citeauthor{Dekel+05} 
are relevant, then how can dark matter-rich spirals make
elliptical-like merger remnants similar to the observed elliptical galaxies,
while kinematical modelling of the PN motions around
ellipticals imply little or no dark matter \citep{Romanowsky+03}?

\begin{figure}
\includegraphics[width=0.85\hsize]{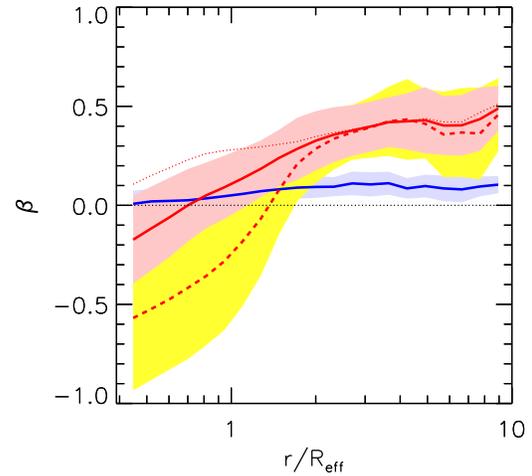}
\caption{
Radial (3D) profiles \citep{Dekel+05} 
of velocity anisotropy, for the dark matter (\emph{nearly
constant solid}), young
stars (\emph{dashed}), old stars (\emph{dotted}) and all stars
(\emph{rising solid curves}), respectively. 
%From \cite{Dekel+05}.
\label{betasims}}
\end{figure}
There are several reasons for the inconsistency between the kinematical and
dynamical analyses.
First, the orbits of the outer stars turn out to be quite radial,
with $\beta \approx 0.5$ (Fig.~\ref{betasims} from \citeauthor{Dekel+05} and
also 
\citealp{DSS04,ANS05}),  
which decreases the
observed velocity dispersions, since the dispersion 
at a fixed projected radius
depends \citep{ML05b} on the mass profile outside that radius and the overall
anisotropy profile. 
This radial stellar anisotropy arises because the stars selected to lie far
from the center in the remnant must have come from the inner regions right
after the closest approach, and were thrown out on elongated orbits in tidal
tails (see Fig.~\ref{snapshots}).
\begin{figure}
  \includegraphics[width=\hsize]{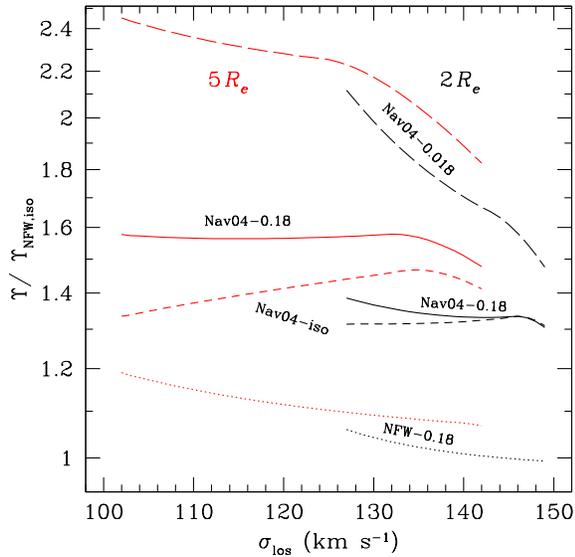}
  \caption{Mass-to-light ratio at the
virial radius, relative to that derived from isotropic NFW modelling, as a
  function of observed line-of-sight stellar velocity dispersion at 2
  (\emph{right}) 
  and 5 (\emph{left})$\,R_e$.
From \cite{ML05b}.
}
\label{upsbias}
\end{figure}
This is illustrated in
Fig.~\ref{upsbias}, which shows that for the stellar anisotropy of the merger
simulations (upper curves: `Nav04-0.18') the mass inferred at the virial radius with the
recent representation \citep{Navarro+04} of $\Lambda$CDM density profiles is 2.4
times what would be inferred for an isotropic ``standard'' NFW \citep{NFW96} 
dark matter model \citep{ML05b}.
Although the best fit models found in the kinematical analysis are moderately
radially anisotropic (Romanowsky 2005, private communication), our equal mass
merger leads to even stronger radial anisotropy, and unequal mass major mergers
lead to even stronger anisotropy.

Second, the galaxy modelled in detail by \citeauthor{Romanowsky+03}, NGC~3379, 
appears circular and its PN system shows no rotation
(\citeauthor{CJD93,Romanowsky+03})
and may well be an oblate elliptical viewed face-on \citep{SS99}, 
which would lead to roughly 10\% lower velocity 
dispersions (\citeauthor{Dekel+05}).

\begin{figure}
  \includegraphics[width=\hsize]{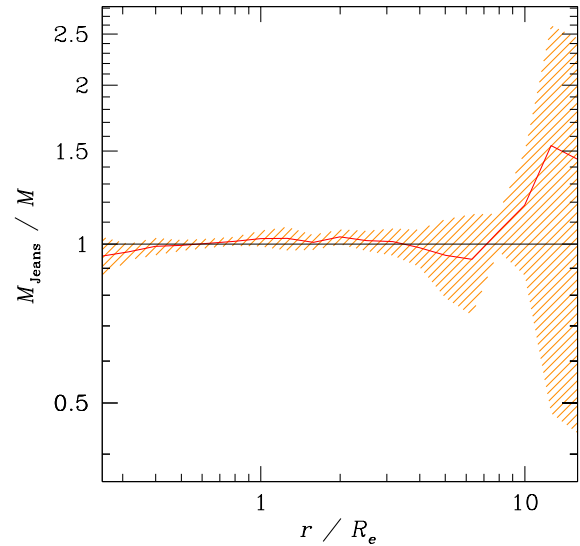}
  \caption{Ratio of Jeans mass (i.e., the mass from the stationary spherical
  Jeans equation [\ref{jeanseq}] --- with the velocity dispersions replaced
  by rms velocities to take streaming motions into account), 
to true mass averaged over the 10 simulated merger
  remnants, and in 
  spherical shells.}
\label{mjeans}
\end{figure}

Moreover, the galaxy may not be stationary, contrary to the assumptions of all
kinematical modelling. However, Fig.~\ref{mjeans} indicates that the
spherical stationary Jeans
equation (\ref{jeanseq}) 
recovers well the true mass profile, so that 
% for all purposes 
the merger remnants can be modelled as 
stationary and
spherical systems (at the meeting, we
incorrectly reported
a 40\% deficit in the mass estimated by the stationary spherical 
Jeans equation).

\section{Are the PNe observed in elliptical galaxies young?}

\begin{figure}
  \includegraphics[width=\hsize]{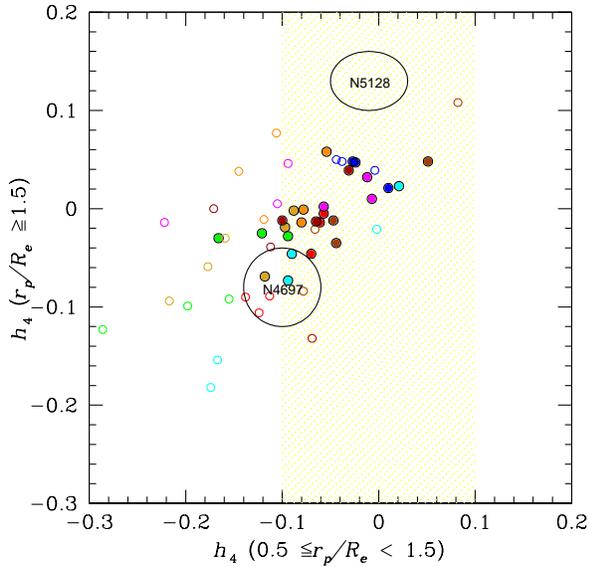}
  \caption{Outer vs. inner Gauss-Hermite $h_4$ coefficient for simulated 
young stars and for PNe in observed galaxies (the ellipses represent
  confidence intervals obtained using bootstraps). The shaded region
  indicates the 
  $h_4$ values obtained from spectral absorption lines of elliptical
  galaxies.
%From \citeauthor{Dekel+05} (supplementary material).
}
\label{h4stats}
\end{figure}

There are several reasons to believe that the
PNe observed in elliptical galaxies are young
objects, although each explanation comes with its caveat:
\begin{enumerate}
\item in the latest model of PN evolution \citep{Marigo+04}, the brightest PNe
are 1 or so Gyr old, comparable to our young stars --- however, the
universality of the bright-end of the PN luminosity function \citep{CJF89},
between old ellipticals and young spirals,
suggests otherwise (see below);
\item the young stars in the simulations 
% analyzed by \citeauthor{Dekel+05} 
formed during the merger from the 
gas particles, have lower velocity dispersions than the older stars (dashed
vs. solid curves in Fig.~\ref{siglossims}) --- however, the effect is of
order of 10\%, and the old stars fit the observed PNe velocity dispersions as
well as the young stars;
\item the reduced kurtoses, $h_4$, of the line-of-sight velocity distribution
of the young stars are larger than for the
old stars and more similar to the $h_4$ parameters measured in the two
galaxies with publicly available PN velocity data, while the $h_4$ values for
the old stars are closer to the measurements from spectral absorption lines
(see Fig.~\ref{h4stats}) --- however, the match is not excellent.
\end{enumerate}

None of these reasons is compelling. One clever reconciliation (\citealp{CSFJ05}
and Ciardullo, in
these proceedings) between the observed universality of the PNLF between
young spirals and old elliptical galaxies and the recent \citeauthor{Marigo+04}
model linking the bright-end PNLF with the age of the PNe 
is to suppose that the progenitors of the bright PNe in ellipticals are
binary systems (blue stragglers) of fairly old 
stellar mass stars rather than much 
younger $2\,M_\odot$ single stars.

Alternatively, it may be possible to reconciliate PNLF universality and the
\citeauthor{Marigo+04} PN model if one supposes that the bright PNe
in ellipticals are caused by recent (1 Gyr) accretion of very low mass
galaxies that go through a star burst as they they are tidally squeezed by
the gravitational potential of the elliptical. Contrary to any model driven
by major mergers, for which 1 Gyr old starbursts will be as stochastic as is
the merging process, the rate of accretion should vary little with galaxy
type, luminosity and perhaps even environment.

Both scenarios ought to be tested quantitatively in the context of
evolutionary models for the PN luminosities, stellar masses and galaxy spectra.
% \citeauthor{Marigo+04} model, as well as models for the evolution of the
% stellar mass function.
The mass accretion rate can be obtained
\citep{SSM98} through the extended Press-Schechter formalism
(e.g. \citealp{LC93}). 
Will there be sufficient blue
stragglers in ellipticals to produce the universality of the PNLF? And, on
the other hand, 
will the rate of accretion of small galaxies be large enough to produce
sufficient bright PNe and not too large to make elliptical galaxies bluer
than they are observed to be?

\begin{theacknowledgments}
We thank Ryszard Szczerba and Grazyna Stasi\'nska for helping
GAM attend this superbly organized meeting and another outstanding
Polish astronomer, Ewa {\L}okas, who collaborated on work that led to
Fig.~\ref{upsbias}, as well as T.~J. Cox for his beautiful simulations.
%   Infandum, regina, iubes renovare dolorem, Troianas ut opes et
%   lamentabile regnum cruerint Danai; quaeque ipse miserrima vidi, et
%   quorum pars magna fui. Quis talia fando Myrmidonum Dolopumve aut duri
%   miles Ulixi temperet a lacrimis?
\end{theacknowledgments}



\bibliographystyle{aipproc}   % if natbib is available

\bibliography{master}

\begin{thebibliography}{20}
\expandafter\ifx\csname natexlab\endcsname\relax\def\natexlab#1{#1}\fi
\providecommand{\enquote}[1]{``#1''}
\expandafter\ifx\csname url\endcsname\relax
  \def\url#1{\texttt{#1}}\fi
\expandafter\ifx\csname urlprefix\endcsname\relax\def\urlprefix{URL }\fi
\providecommand{\eprint}[2][]{\url{#2}}

\bibitem[{Ciardullo} et~al.(1993)]{CJD93}
R.~{Ciardullo}, G.~H. {Jacoby}, and H.~B. {Dejonghe}, \emph{\apj},
  \textbf{414}, 454--462 (1993).

\bibitem[{M{\' e}ndez} et~al.(2001)]{Mendez+01}
R.~H. {M{\' e}ndez}, A.~{Riffeser}, R.-P. {Kudritzki}, M.~{Matthias}, K.~C.
  {Freeman}, M.~{Arnaboldi}, M.~{Capaccioli}, and O.~E. {Gerhard}, \emph{\apj},
  \textbf{563}, 135--150 (2001).

\bibitem[{Romanowsky} et~al.(2003)]{Romanowsky+03}
A.~J. {Romanowsky}, N.~G. {Douglas}, M.~{Arnaboldi}, K.~{Kuijken}, M.~R.
  {Merrifield}, N.~R. {Napolitano}, M.~{Capaccioli}, and K.~C. {Freeman},
  \emph{Science}, \textbf{301}, 1696--1698 (2003).

\bibitem[{Springel} et~al.(2001)]{SWTK01}
V.~{Springel}, S.~D.~M. {White}, G.~{Tormen}, and G.~{Kauffmann},
  \emph{\mnras}, \textbf{328}, 726--750 (2001).

\bibitem[{Cox} et~al.(2005)]{CJPS05}
T.~J. {Cox}, P.~{Jonsson}, J.~R. {Primack}, and R.~S. {Somerville},
  \emph{\mnras} (2005), submitted, arXiv:astro-ph/0503201.

\bibitem[{Dekel} et~al.(2005)]{Dekel+05}
A.~{Dekel}, F.~{Stoehr}, G.~A. {Mamon}, T.~J. {Cox}, and J.~R. {Primack},
  \emph{\nat} (2005), in press, arXiv:astro-ph/0501622.

\bibitem[{Jedrzejewski} and {Schechter}(1989)]{JS89}
R.~{Jedrzejewski}, and P.~L. {Schechter}, \emph{\aj}, \textbf{98}, 147--165
  (1989).

\bibitem[{Binney} et~al.(1990)]{BDI90}
J.~J. {Binney}, R.~L. {Davies}, and G.~D. {Illingworth}, \emph{\apj},
  \textbf{361}, 78--97 (1990).

\bibitem[{Bender} et~al.(1994)]{BSG94}
R.~{Bender}, R.~P. {Saglia}, and O.~E. {Gerhard}, \emph{\mnras}, \textbf{269},
  785--813 (1994).

\bibitem[{Statler} and {Smecker-Hane}(1999)]{SS99}
T.~S. {Statler}, and T.~{Smecker-Hane}, \emph{\aj}, \textbf{117}, 839--854
  (1999).

\bibitem[{Dom{\'{\i}}nguez-Tenreiro} et~al.(2004)]{DSS04}
R.~{Dom{\'{\i}}nguez-Tenreiro}, A.~{S{\'a}iz}, and A.~{Serna}, \emph{\apjl},
  \textbf{611}, L5--L8 (2004).

\bibitem[{Abadi} et~al.(2005)]{ANS05}
M.~G. {Abadi}, J.~F. {Navarro}, and M.~{Steinmetz}, \emph{\mnras} (2005),
  submitted, arXiv:astro-ph/0506659.

\bibitem[{Mamon} and {{\L}okas}(2005)]{ML05b}
G.~A. {Mamon}, and E.~L. {{\L}okas}, \emph{\mnras} (2005), in press,
  arXiv:astro-ph/0405491.

\bibitem[{Navarro} et~al.(2004)]{Navarro+04}
J.~F. {Navarro}, E.~{Hayashi}, C.~{Power}, A.~R. {Jenkins}, C.~S. {Frenk},
  S.~D.~M. {White}, V.~{Springel}, J.~{Stadel}, and T.~R. {Quinn},
  \emph{\mnras}, \textbf{349}, 1039--1051 (2004).

\bibitem[{Navarro} et~al.(1996)]{NFW96}
J.~F. {Navarro}, C.~S. {Frenk}, and S.~D.~M. {White}, \emph{\apj},
  \textbf{462}, 563--575 (1996).

\bibitem[{Marigo} et~al.(2004)]{Marigo+04}
P.~{Marigo}, L.~{Girardi}, A.~{Weiss}, M.~A.~T. {Groenewegen}, and C.~{Chiosi},
  \emph{\aap}, \textbf{423}, 995--1015 (2004).

\bibitem[{Ciardullo} et~al.(1989)]{CJF89}
R.~{Ciardullo}, G.~H. {Jacoby}, and H.~C. {Ford}, \emph{\apj}, \textbf{344},
  715--725 (1989).

\bibitem[{Ciardullo} et~al.(2005)]{CSFJ05}
R.~{Ciardullo}, S.~{Sigurdsson}, J.~J. {Feldmeier}, and G.~H. {Jacoby},
  \emph{\apj}, \textbf{629}, 499--506 (2005).

\bibitem[{Salvador-Sol\'e} et~al.(1998)]{SSM98}
E.~{Salvador-Sol\'e}, J.~M. {Solanes}, and A.~{Manrique}, \emph{\apj},
  \textbf{499}, 542--547 (1998).

\bibitem[{Lacey} and {Cole}(1993)]{LC93}
C.~{Lacey}, and S.~{Cole}, \emph{\mnras}, \textbf{262}, 627--649 (1993).

\end{thebibliography}

\IfFileExists{\jobname.bbl}{}
 {\typeout{}
  \typeout{******************************************}
  \typeout{** Please run "bibtex \jobname" to optain}
  \typeout{** the bibliography and then re-run LaTeX}
  \typeout{** twice to fix the references!}
  \typeout{******************************************}
  \typeout{}
 }

\end{document}